\documentclass[twocolumn,prl]{revtex4}
\usepackage{amsfonts}
\usepackage[T1]{fontenc}
\usepackage{amsmath,amsbsy,amssymb,graphicx}
\usepackage{times}

\let\mathbf=\boldsymbol

\begin{document}

\title{{\Large Topological Kirchhoff Law }\\
{\Large and Bulk-Edge Correspondence for Valley-Chern and Spin-Valley-Chern
Numbers }}
\author{Motohiko Ezawa}
\affiliation{Department of Applied Physics, University of Tokyo, Hongo 7-3-1, 113-8656,
Japan }

\begin{abstract}
The valley-Chern and spin-valley-Chern numbers are the key concepts in
valleytronics. They are topological numbers in the Dirac theory but not in
the tight-binding model. We analyze the bulk-edge correspondence between the
two phases which have the same Chern and spin-Chern numbers but different
valley-Chern and spin-valley-Chern numbers. The edge state between them is
topologically trivial in the tight-binding model but is shown to be as
robust as the topological edge. We construct Y-junctions made of topological
edges. They satisfy the topological Kirchhoff law, where the topological
charges are conserved at the junction. We may interpret a Y-junction as a
scattering process of particles which have four topological numbers. It
would be a milestone of future topological electronics.
\end{abstract}

\maketitle


Topological insulator is one of the most fascinating concepts found in this
decade\cite{Hasan,Qi}. It is characterized by topological numbers such as
the Chern ($\mathcal{C}$) number and the $\mathbb{Z}_{2}$ index. When the
spin s$_{z}$ is a good quantum number, the spin-Chern ($\mathcal{C}_{s}$)
number replaces the role of the $\mathbb{Z}_{2}$ index\cite%
{Prodan,Sheng,Yang}. We consider honeycomb lattice systems. Electrons
resides either in the $K$ or $K^{\prime }$ valley in the low-energy Dirac
theory. Accordingly we can define the valley-Chern ($\mathcal{C}_{v}$) number%
\cite{Fang11,Fang13,Li} and the spin-valley-Chern ($\mathcal{C}_{sv}$) number%
\cite{Fang11} in the Dirac theory. This valley degree of freedom leads to
valleytronics\cite%
{Rycerz,Qiao10,Qiao11,Qiao12,Qiao13,Xiao07,Xiao12,Cao,Jung,Yao,Tse,Ding}.
However, the $\mathcal{C}_{v}$ and $\mathcal{C}_{sv}$ numbers are
ill-defined in the tight-binding model because the topological numbers are
defined by the summation of Berry curvatures over the entire Brillouin zone.
Namely, a state is indexed by the two topological numbers in the
tight-binding model, while it is indexed by the four topological numbers in
the Dirac theory.

The are four independent spin-valley dependent Chern numbers in the Dirac
theory of honeycomb systems. Each Chern number can be controlled
independently by changing the sign of spin-valley dependent Dirac masses.
There are 16 types of topological insulators, as shown in the table \ref%
{tableA}. They are quantum anomalous Hall (QAH) insulator, four types of
spin-polarized QAH (SQAH) insulators, quantum spin Hall (QSH) insulator and
the band insulator with charge-density-wave (CDW) or antiferromagnetic (AF)
order. The CDW and AF insulators are regarded trivial in the tight-binding
model.

\begin{table}[tbp]
\begin{center}
\begin{tabular}{|c|c|c|c|c|c|c|c|c|}
\hline
& $C_{\uparrow }^{K}$ & $C_{\uparrow }^{K^{\prime }}$ & $C_{\downarrow }^{K}$
& $C_{\downarrow }^{K^{\prime }}$ & $C$ & $2C_{s}$ & $C_{v}$ & $2C_{sv}$ \\ 
\hline
QAH & $1/2$ & $1/2$ & $1/2$ & $1/2$ & $2$ & $0$ & $0$ & $0$ \\ \hline
SQAH$_{1}$ & $1/2$ & $1/2$ & $1/2$ & $-1/2$ & $1$ & $1$ & $1$ & $-1$ \\ 
\hline
SQAH$_{2}$ & $1/2$ & $-1/2$ & $1/2$ & $1/2$ & $1$ & $1$ & $-1$ & $1$ \\ 
\hline
SQAH$_{3}$ & $1/2$ & $1/2$ & $-1/2$ & $1/2$ & $1$ & $-1$ & $1$ & $1$ \\ 
\hline
SQAH$_{4}$ & $-1/2$ & $1/2$ & $1/2$ & $1/2$ & $1$ & $-1$ & $-1$ & $-1$ \\ 
\hline
QSH & $1/2$ & $-1/2$ & $1/2$ & $-1/2$ & $0$ & $2$ & $0$ & $0$ \\ \hline
CDW & $1/2$ & $1/2$ & $-1/2$ & $-1/2$ & $0$ & $0$ & $2$ & $0$ \\ \hline
AF & $1/2$ & $-1/2$ & $-1/2$ & $1/2$ & $0$ & $0$ & $0$ & $2$ \\ \hline
\end{tabular}%
\end{center}
\caption{Corresponding to the spin and valley degrees of freedom, there are $%
4$ Chern numbers $\mathcal{C}_{s_{z}}^{\protect\eta }$, each of which takes $%
\pm \frac{1}{2}$. Equivalently they are given by the Chern ($\mathcal{C}$),
spin-Chern ($\mathcal{C}_{s}$), valley-Chern ($\mathcal{C}_{v}$) and
spin-valley-Chern ($\mathcal{C}_{sv}$) numbers. Hence there are $16$ states
indexed by them, among which $8$ states are explicitly displayed. The other $%
8$ states are the conjugate states such as QAH$^{\ast }$ whose topolorical
numbers are given by $-\mathcal{C}_{s_{z}}^{\protect\eta }$.}
\label{tableA}
\end{table}

In this paper, we study the bulk-edge correspondence with respect to the $%
\mathcal{C}_{v}$ and $\mathcal{C}_{sv}$ numbers by examining the boundary of
two insulators which have the same $\mathcal{C}$ and $\mathcal{C}_{s}$
numbers but different $\mathcal{C}_{v}$ and $\mathcal{C}_{sv}$ numbers.
First we show that gapless edge states appear though they are trivial in the
tight-binding model. Furthermore, we show that they are as robust as the
topologically protected edges.

We propose a topological electronics based on the edge states in the Dirac
theory. We are able to assign four topological numbers to each edge states.
By joining three different topological insulators at one point, we can
construct a Y-junction made of topological edge states. The edge states at
the junction satisfies the conservation of four topological numbers, which
we call the topological Kirchhoff law. We can change the connectivity of
edge states by changing the topological property of bulk insulators, for
instance, by applying electric field. The process may be interpreted as a
pair annihilation of two Y-junctions.

\textbf{Hamiltonian:} The honeycomb lattice consists of two sublattices made
of $A$ and $B$ sites. We consider a buckled system with the layer separation 
$2\ell $ between these two sublattices. The states near the Fermi energy are 
$\pi $ orbitals residing near the $K$ and $K^{\prime }$ points at opposite
corners of the hexagonal Brillouin zone. The low-energy dynamics in the $K$
and $K^{\prime }$ valleys is described by the Dirac theory. In what follows
we use notations $s_{z}=\uparrow \downarrow $, $t_{z}=A,B$, $\eta
=K,K^{\prime }$ in indices while $s_{z}^{\alpha }=\pm 1$ for $\alpha
=\uparrow \downarrow $, $t_{z}^{i}=\pm 1$ for $i=A$,$B$, and $\eta _{i}=\pm
1 $ for $i=K,K^{\prime }$ in equations. We also use the Pauli matrices $%
\sigma _{a}$ and $\tau _{a}$ for the spin and the sublattice pseudospin,
respectively.

We have previously proposed a generic Hamiltonian for honeycomb systems\cite%
{Ezawa2Ferro}, which contains eight interaction terms mutually commutative
in the Dirac limit. Among them four contribute to the Dirac mass. The other
four contribute to the shift of the energy spectrum. We are able to make a
full control of the Dirac mass and the energy shift independently at each
spin and valley by varying these parameters, and materialize various
topological phases\cite{EzawaQAHE,EzawaPhoto}.

By taking those affecting the Dirac mass, the tight-binding model is given by%
\cite{KaneMele,LiuPRB,Ezawa2Ferro}, 
\begin{eqnarray}
H &=&-t\sum_{\left\langle i,j\right\rangle \alpha }c_{i\alpha }^{\dagger
}c_{j\alpha }+i\frac{\lambda _{\text{SO}}}{3\sqrt{3}}\sum_{\left\langle
\!\left\langle i,j\right\rangle \!\right\rangle \alpha \beta }\nu
_{ij}c_{i\alpha }^{\dagger }\sigma _{\alpha \beta }^{z}c_{j\beta }  \notag \\
&&-\lambda _{V}\sum_{i\alpha }\mu _{i}c_{i\alpha }^{\dagger }c_{i\alpha }+i%
\frac{\lambda _{\Omega }}{3\sqrt{3}}\sum_{\left\langle \!\left\langle
i,j\right\rangle \!\right\rangle \alpha \beta }\nu _{ij}c_{i\alpha
}^{\dagger }c_{j\beta }  \notag \\
&&+\lambda _{\text{SX}}\sum_{i\alpha }\mu _{i}c_{i\alpha }^{\dagger }\sigma
_{\alpha \alpha }^{z}c_{i\alpha },
\end{eqnarray}%
where $c_{i\alpha }^{\dagger }$ creates an electron with spin polarization $%
\alpha $ at site $i$, and $\left\langle i,j\right\rangle /\left\langle
\!\left\langle i,j\right\rangle \!\right\rangle $ run over all the
nearest/next-nearest neighbor hopping sites. We explain each term. The first
term represents the nearest-neighbor hopping with the transfer energy $t$.
The second term represents the SO coupling\cite{KaneMele} with $\lambda _{%
\text{SO}}$. The third term is the staggered sublattice potential term\cite%
{EzawaNJP} with $\lambda _{V}=\ell E_{z}$ in electric field $E_{z}$. The
forth term is the Haldane term\cite{Haldane} with $\lambda _{\Omega }$. The
fifth term represents the antiferromagnetic exchange magnetization\cite%
{Ezawa2Ferro,Feng} with $\lambda _{\text{SX}}$.

We give typical sample parameters though we treat them as free parameters.
Silicene is a good candidate, where $t=1.6$eV, $\lambda _{\text{SO}}=3.9$meV
and $\ell =0.23$\AA . The Haldane term might be induced by the
photo-irradiation, where $\lambda _{\Omega }=\hbar v_{\text{F}}^{2}\mathcal{A%
}^{2}\Omega ^{-1}$ with $\Omega $ the frequency and $\mathcal{A}$ the
dimensionless intensity\cite{EzawaPhoto,Kitagawa01B,Oka}. The
antiferromagnetic exchange magnetization might be induced by certain
proximity effects. The second candidate is provskite transition-metal-oxide
grown on [111]-direction, which has antiferromagnetic order intrinsically%
\cite{Hu}. This material has also the buckled structure as in the case of
silicene. Parameters are $t\approx 0.2$eV, $\lambda _{\text{SO}}=7.3$meV, $%
\lambda _{V}=\ell E_{z}$, $\lambda _{\text{SX}}=141$meV for LaCrAgO.

The low-energy Hamiltonian is described by\cite{Ezawa2Ferro} 
\begin{eqnarray}
H_{\eta } &=&\hbar v_{\text{F}}\left( \eta k_{x}\tau _{x}+k_{y}\tau
_{y}\right) +\lambda _{\text{SO}}\sigma _{z}\eta \tau _{z}  \notag \\
&&-\lambda _{V}\tau _{z}+\lambda _{\Omega }\eta \tau _{z}+\lambda _{\text{SX}%
}\sigma _{z}\tau _{z},
\end{eqnarray}%
where $v_{\text{F}}=\frac{\sqrt{3}}{2\hbar }at$ is the Fermi velocity. The
coefficient of $\tau _{z}$ is the mass of Dirac fermions in the Hamiltonian,%
\begin{equation}
\Delta _{s_{z}}^{\eta }=\eta s_{z}\lambda _{\text{SO}}-\lambda _{V}+\eta
\lambda _{\Omega }+s_{z}\lambda _{\text{SX}}.  \label{DiracMass}
\end{equation}%
The band gap is given by $2|\Delta _{s_{z}}^{\eta }|$.

\textbf{Topological numbers:} We consider the systems where the spin $s_{z}$
is a good quantum number. The summation of the Berry curvature over all
occupied states of electrons with spin $s_{z}$ in the Dirac valley $K_{\eta
} $ yields\cite{Hasan,Qi,Diamag}%
\begin{equation}
\mathcal{C}_{s_{z}}^{\eta }={\frac{\eta }{2}}\text{sgn}(\Delta
_{s_{z}}^{\eta }).  \label{ChernNumbe}
\end{equation}%
There are four independent spin-valley dependent Dirac masses determined by
the four parameters $\lambda _{\text{SO}}$, $\lambda _{V}$, $\lambda
_{\Omega }$ and $\lambda _{\text{SX}}$. Accordingly, we can define 
\begin{eqnarray}
\mathcal{C} &=&\mathcal{C}_{\uparrow }^{\text{K}}+\mathcal{C}_{\uparrow }^{%
\text{K'}}+\mathcal{C}_{\downarrow }^{\text{K}}+\mathcal{C}_{\downarrow }^{%
\text{K'}}, \\
\mathcal{C}_{s} &=&\frac{1}{2}(\mathcal{C}_{\uparrow }^{\text{K}}+\mathcal{C}%
_{\uparrow }^{\text{K'}}-\mathcal{C}_{\downarrow }^{\text{K}}-\mathcal{C}%
_{\downarrow }^{\text{K'}}),
\end{eqnarray}%
and%
\begin{eqnarray}
\mathcal{C}_{v} &=&\mathcal{C}_{\uparrow }^{\text{K}}-\mathcal{C}_{\uparrow
}^{\text{K'}}+\mathcal{C}_{\downarrow }^{\text{K}}-\mathcal{C}_{\downarrow
}^{\text{K'}}, \\
\mathcal{C}_{sv} &=&\frac{1}{2}(\mathcal{C}_{\uparrow }^{\text{K}}-\mathcal{C%
}_{\uparrow }^{\text{K'}}-\mathcal{C}_{\downarrow }^{\text{K}}+\mathcal{C}%
_{\downarrow }^{\text{K'}}).
\end{eqnarray}%
It is to be emphasized that $\mathcal{C}_{v}$ and $\mathcal{C}_{sv}$ are not
defined in the tight-binding model.

The possible sets of topological numbers are $(0,0)$, $(2,0)$, $(0,1)$, $(1,%
\frac{1}{2})$, $(1,-\frac{1}{2})$ up to the overall sign $\pm $ in the
tight-binding model. They are the trivial, QAH, QSH and two types of SQAH
insulators, respectively. They are further classified into subsets according
to the valley degree of freedom in the Dirac theory. Trivial insulators are
divided into two; one with CDW order and the other with AF orders\cite%
{Ezawa2Ferro}. Each type of SQAH insulators are divided into two: There are
four types in all, which we denote by SQAH$_{i}$ with $i=1,2,3,4$. All of
them are summarized in Table \ref{tableA}.

\begin{figure}[t]
\centerline{\includegraphics[width=0.5\textwidth]{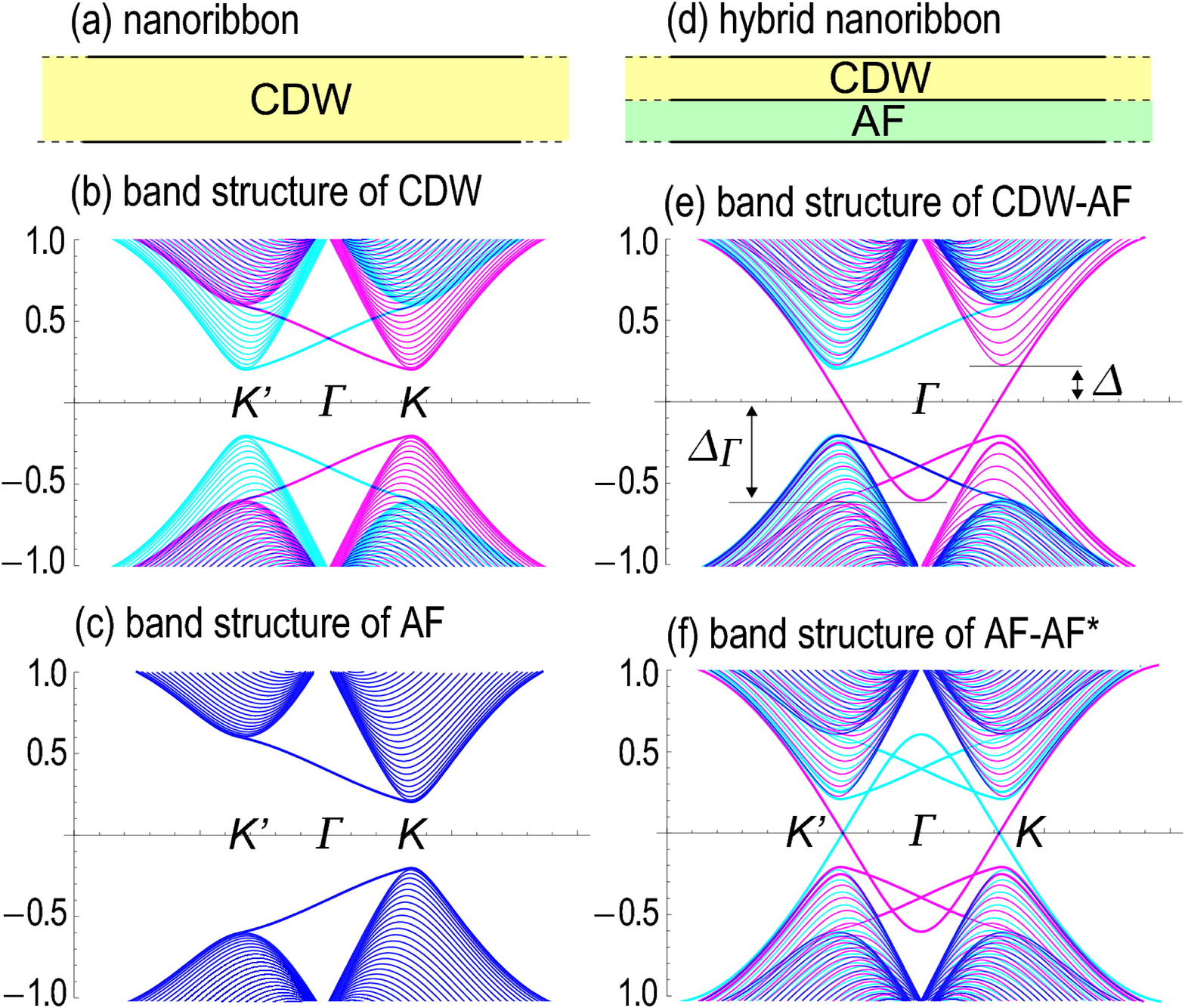}}
\caption{(Color online) Illustration of (a) a nanoribbon and (d) a hybrid
nanoribbon. Band structure of nanoribbon made of (b) CDW and (c) AF
insulators. Band structure of a hybrid nanoribbon made of (e) CDW-AF, and
(f) AF-AF$^{\ast }$ insulators. Up(down)-spin bands are shown in magenta
(cyan). We have taken $\protect\lambda _{\text{SO}}=0.2t$, $\protect\lambda %
_{V}=0.4t$ and $\protect\lambda _{\text{SX}}=0.4t$ unless zero for
illustration.}
\label{FigValleyRibbon}
\end{figure}

\textbf{Bulk-edge correspondence:} The most convenient way to determine the
topological charges is to employ the bulk-edge correspondence. When there
are two topological distinct phases, a topological phase transition may
occur between them. It is generally accepted that the band gap must close at
the topological phase transition point since the topological number cannot
change its quantized value without gap closing. Note that the topological
number is only defined in the gapped system and remains unchanged for any
adiabatic process. Alternatively, we may consider a junction separating two
different topological phases in a single honeycomb system\cite{EzawaNJP}.
Gapless edge modes must appear along the boundary. We may as well analyze
the energy spectrum of a nanoribbon in a topological phase, because the
boundary of the nanoribbon separates a topological state and the vacuum
whose topological numbers are zero: See Fig.\ref{FigValleyRibbon}(a).

\textbf{CDW-AF junction:} We first investigate the trivial insulator in the
tight-binding model, which consists of two subsets (CDW and AF) in the Dirac
theory. It is well known that a nanoribbon made of either the CDW insulator
or the AF insulator has no gapless edge modes, as is regarded to be a
demonstration of their triviality: See Fig.\ref{FigValleyRibbon}(b) and (c).
One may wonder how they can be topological in the Dirac theory without
gapless edge modes in view of the bulk-edge correspondence. The answer to
this problem is that the $\mathcal{C}_{v}$ and $\mathcal{C}_{sv}$ numbers
are not defined in the vacuum. Indeed, only the charge and the spin are well
defined to be zero in the vacuum. Note that the gap needs not close at such
a boundary because $\mathcal{C}_{v}$ and $\mathcal{C}_{sv}$ are defined only
inside of the nanoribbon. This explains the absence of gapless edge modes in
Fig.\ref{FigValleyRibbon}(b) and (c).

We investigate the junction made of the CDW and AF insulators, whose
topological numbers are $(\mathcal{C},\mathcal{C}_{s},\mathcal{C}_{v},%
\mathcal{C}_{sv})=(0,0,2,0)$ and $(0,0,0,2)$, respectively. On one hand, we
expect no gapless edge modes in the tight-binding model. On the other hands,
there should be gapless edge modes in the Dirac theory. We ask how these two
properties are compatible.

To answer this problem we study a hybrid nanoribbon by separating a
nanoribbon into two parts, one in the CDW phase and the other in the AF
phase: See Fig.\ref{FigValleyRibbon}(d). Only the $\mathcal{C}_{v}$ and $%
\mathcal{C}_{sv}$ numbers change across the boundary separating these two
regions. We have calculated the band structure of such a hybrid nanoribbon,
whose result we display in Fig.\ref{FigValleyRibbon}(e). We find one edge
state crossing the Fermi energy twice. It is a manifestation of the fact
that the $\mathcal{C}$ and $\mathcal{C}_{s}$ numbers do not change. On the
other hand, when we concentrate on the vicinity of the $K$ and $K^{\prime }$
points, there are well defined edge states. It is a manifestation of the
fact that the $\mathcal{C}_{v}$ and $\mathcal{C}_{sv}$ numbers change at the
junction.

We proceed to argue how strongly the $\mathcal{C}_{v}$ and $\mathcal{C}_{sv}$
numbers are protected. The word "topologically protected" means that the
edge states are robust against perturbations whose magnitude is less the
bulk gap $\Delta $. Indeed, the bulk gap may close by perturbations stronger
than it, invalidating the topological arguments at all. How robust is the
edge mode at the CDW-AF junction? We have checked that the edge mode takes
the extremal energy $\Delta _{\Gamma }$ at the $\Gamma $ point with%
\begin{equation}
|\Delta _{\Gamma }|=t-\lambda _{V}-\lambda _{\text{SX}}.  \label{GammaGap}
\end{equation}%
It costs the energy $\Delta _{\Gamma }$ to remove the edge states. Now, the
topological edge states are protected by the bulk energy gap 
\begin{equation}
\Delta =\left\vert \eta s_{z}\lambda _{\text{SO}}-\lambda _{V}-s_{z}\lambda
_{\text{SX}}\right\vert .  \label{Gap}
\end{equation}%
We find $\Delta _{\Gamma }\gg \Delta $ since $\Delta _{\Gamma }$ is the
order of eV, while $\Delta $ is the order of meV. It is very robust against
perturbations. 
\begin{figure}[t]
\centerline{\includegraphics[width=0.45\textwidth]{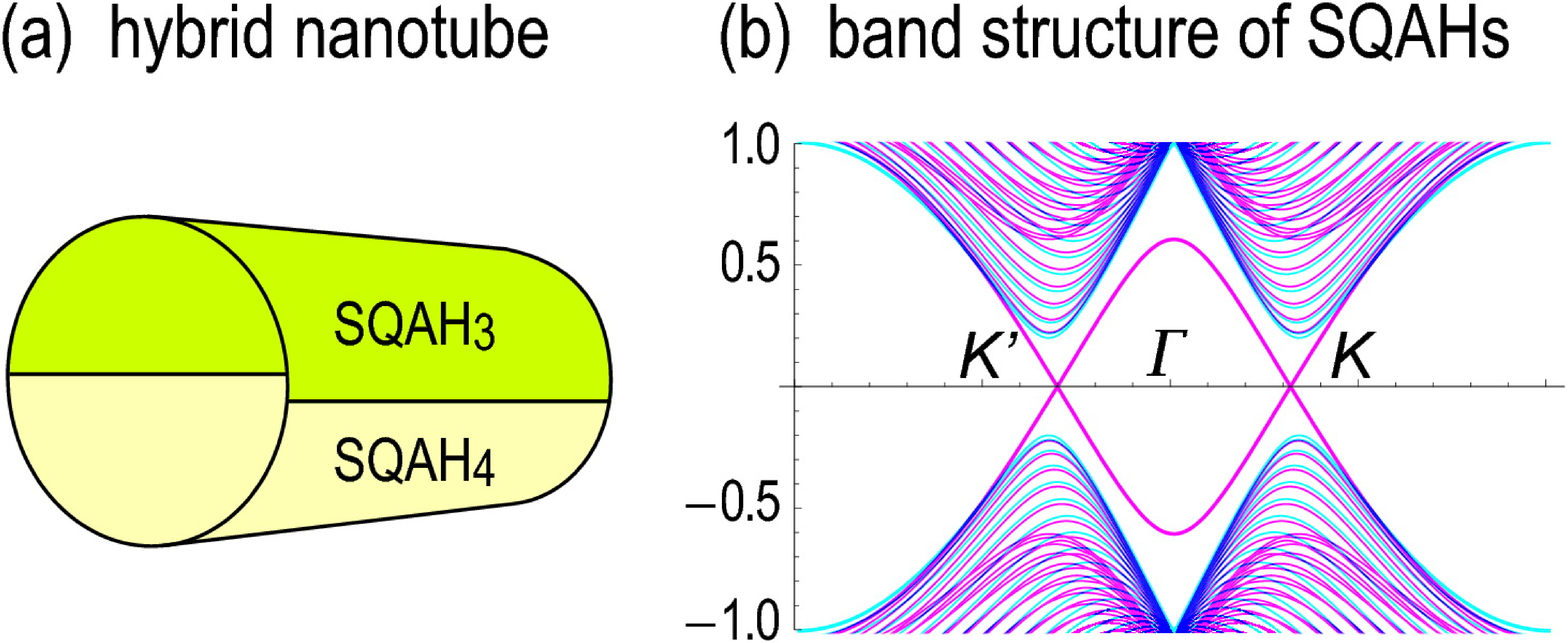}}
\caption{(Color online) (a) Illustration of a hybrid nanotube. (b) Band
structure of a hybrid nanotube made of SQAH$_{1}$ and SQAH$_{2}$.
Up(down)-spin bands are shown in magenta (cyan). We have taken $\protect%
\lambda _{\text{SO}}=\protect\lambda _{V}=\protect\lambda _{\text{SX}}=0.2t$
for illustration.}
\label{FigSQAH}
\end{figure}

\textbf{AF-AF}$^{\ast }$\textbf{\ junction:} The second example is given by
the junction made of AF with $(0,0,0,2)$ and AF$^{\ast }$ with $(0,0,0,-2)$.
The phase boundary is an antiferromagnetic domain wall with the
magnetization reversed at a line defect. Let us take the line along the $x$
axis. The junction is created by introducing the order parameter such that $%
\lambda _{\text{SX}}(y)=\lambda _{\text{SX}}$ for $y>0$ and $-\lambda _{%
\text{SX}}$ for $y<0$. To investigate the edge state located at $y=0$, we
calculate the band structure of a hybrid nanoribbon composed of the AF phase
and the AF$^{\ast }$ phase. We present the result in Fig.\ref%
{FigValleyRibbon}(f). We see clearly gapless edge modes highly enhanced at
the $\Gamma $ point. The extremal energy $\Delta _{\Gamma }$ of the gapless
edge mode is given by (\ref{GammaGap}), while the bulk gap is given by (\ref%
{Gap}) also in this case. The edge is very robust.

\textbf{SQAH-SQAH junction:} We next investigate the junction made of two
different SQAHs. As an explicit example we take SQAH$_{1}$ with $(\mathcal{C}%
,\mathcal{C}_{s},\mathcal{C}_{v},\mathcal{C}_{sv})=(1,1,1,-1)$ and SQAH$_{2}$
with $(1,1,-1,1)$. It is not appropriate to use a hybrid nanoribbon in the
present case since gapless edge modes appear even for a simple nanoribbon in
the SQAH phase. We calculate the band structure of a nanotube geometry since
no gapless edge modes appear even for a simple nanotube in the SQAH phase
owing to the lack of the edge itself. We take a hybrid nanotube where one
half of the nanotube is SQAH$_{1}$ and the other half is SQAH$_{2}$, as
illustrated in Fig.\ref{FigSQAH}(a). We show the result in Fig.\ref{FigSQAH}%
(b), where we see clearly a gapless edge mode highly enhanced at the $\Gamma 
$ point. The extremal energy $\Delta _{\Gamma }$ of the gapless edge mode is
given by (\ref{GammaGap}), while the bulk gap is given by (\ref{Gap}) also
in this case. The edge is very robust as well.

\textbf{Gapless edge mode in Dirac theory: }We proceed to construct the
Dirac theory of the gapless edge states\cite{EzawaNJP}. They emerge along a
curve where the Dirac mass vanishes, $\Delta _{s_{z}}^{\eta }\left(
x,y\right) =0$. Let us take the edge along the $x$ axis. The zero modes
emerge along the line determined by $\Delta _{s_{z}}^{\eta }\left( y\right)
=0$, when $\Delta _{s_{z}}^{\eta }\left( y\right) $ changes the sign. We may
set $k_{x}=$constant due to the translational invariance along the $x$ axis.
We seek the zero-energy solution by setting $\psi _{B}=i\xi \psi _{A}$ with $%
\xi =\pm 1$. Here, $\psi _{A}$ is a two-component amplitude with the up spin
and down spin, $\psi _{A}=(\psi _{A}^{\uparrow },\psi _{A}^{\downarrow })$.
Setting $\psi _{A}\left( x,y\right) =e^{ik_{x}x}\phi _{A}\left( y\right) $,
we obtain $H_{\eta }\psi _{A}\left( x,y\right) =E_{\eta \xi }\psi _{A}\left(
x,y\right) $, together with a linear dispersion relation $E_{\eta \xi }=\eta
\xi \hbar v_{\text{F}}k_{x}$. We can explicitly solve this as 
\begin{equation}
\phi _{A}^{s_{z}}\left( y\right) =C\exp \left[ \frac{\xi }{\hbar v_{\text{F}}%
}\int^{y}\Delta _{s_{z}}^{\eta }\left( y^{\prime }\right) dy^{\prime }\right]
,  \label{ZeroModeSolut}
\end{equation}%
where $C$ is the normalization constant. The sign $\xi $ is determined so as
to make the wave function finite in the limit $\left\vert y\right\vert
\rightarrow \infty $. This is a reminiscence of the Jackiw-Rebbi mode\cite%
{Jakiw} presented for the chiral mode. The difference is the presence of the
spin and valley indices in the wave function.

\begin{figure}[t]
\centerline{\includegraphics[width=0.5\textwidth]{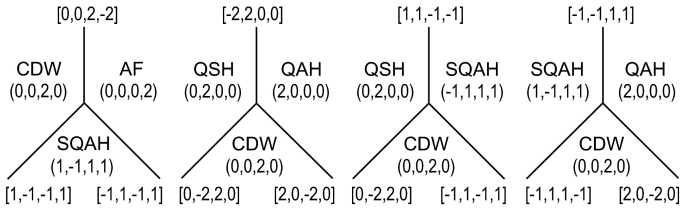}}
\caption{Illustrations of typical Y-junctions. The edge between two
different topological insulators carries a set of four topological charges.
Three edges can make a Y-junction provided that the sum of their topological
charges is zero.}
\label{FigYJunc}
\end{figure}

\textbf{Topological Kirchhoff law:} We consider a configuration where three
different topological insulators meet at one point: See Fig.\ref{FigYJunc}.
In this configuration there are three edges forming a Y-junction. It is
convenient to assign the topological numbers to each edge which are the
difference between those of the two adjacent topological insulators. Namely,
when the topological insulator with $(\mathcal{C}^{L},\mathcal{C}_{s}^{L},%
\mathcal{C}_{v}^{L},\mathcal{C}_{sv}^{L})$ is on the left-hand side of the
one with $(\mathcal{C}^{R},\mathcal{C}_{s}^{R},\mathcal{C}_{v}^{R},\mathcal{C%
}_{sv}^{R})$, we assign the numbers $[\mathcal{C}^{L}-\mathcal{C}^{R},%
\mathcal{C}_{s}^{L}-\mathcal{C}_{s}^{R},\mathcal{C}_{v}^{L}-\mathcal{C}%
_{v}^{R},\mathcal{C}_{sv}^{L}-\mathcal{C}_{sv}^{R}]$ to the boundary, as
illustrated in Fig.\ref{FigYJunc}. The condition which edges can make a
Y-junction is the conservation of these topological numbers at the junction.
This law is a reminiscence of the Kirchhoff law, which dictates the
conservation of currents at the junction of electronic circuits. We call it
the topological Kirchhoff law.

The number of Y-junctions is given by the combination of selecting $3$ from $%
16$ topological insulators, i.e., $_{16}C_{3}=560$. The number of
topological edge states is determined by the combination of selecting $2$
from $16$ topological insulators. We have $_{16}C_{2}=120$ types of
topological edge states. We show typical examples of Y-junctions in Fig.\ref%
{FigYJunc}.

We present an interesting interpretation of the topological Kirchhoff law.
We may regard each topological edge state as a world line of a particle
carrying the four topological charges. The Y-junction may be interpreted as
a scattering process of these particles. In this scattering process, the
topological charges conserve. 
\begin{figure}[t]
\centerline{\includegraphics[width=0.34\textwidth]{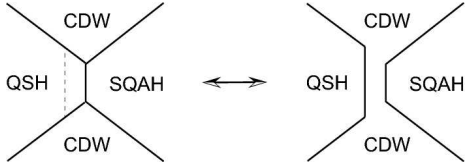}}
\caption{Schematic illustration of topological electronic circuit. The CDW
phase is created by applying electric field $E_{z}$ to the QSH phase beyond
the critical value $E_{\text{cr}}$. Tuning it locally we can change the form
of circuite by a pair annihilatin of two Y-junctions.}
\label{Circuit}
\end{figure}

\textbf{Topological electronic circuits:} We can construct electronic
circuits made of edge states by joining Y-junctions. Each topological edge
state carries conductance\cite{EzawaConduc}, whose magnitude is given by the
Chern number $\mathcal{C}$ in unit of $e^{2}/h$. In general, the edge states
carry charge $\mathcal{C}$, spin $\mathcal{C}_{s}$, valley-charge $\mathcal{C%
}_{v}$ and spin-valley-charge $\mathcal{C}_{sv}$. The present-day electronic
circuits only use the charge degree of freedom. In our circuits of
topological edges we can make a full use of four types of charges. This
would greatly enhance the ability of information processing.

We can control the position of edge state by controlling the parameters of
the bulk states. The easiest way is to apply electric field $E_{z}$ locally.
Let us review the topological phase transition taking place as $E_{z}$
changes\cite{EzawaNJP} by taking $\lambda _{\Omega }=\lambda _{\text{SX}}=0$%
, where the Dirac mass is given by $\Delta _{s_{z}}^{\eta }=\eta
s_{z}\lambda _{\text{SO}}-\ell E_{z}$. The condition $\Delta _{s_{z}}^{\eta
}=0$ implies $E_{z}=\pm E_{\text{cr}}$ with $E_{\text{cr}}=\lambda _{\text{SO%
}}/\ell $. It follows that $(\mathcal{C},\mathcal{C}_{s})=(0,0)$ for $%
|E_{z}|<E_{\text{cr}}$ and $(0,\frac{1}{2})$ for $|E_{z}|>E_{\text{cr}}$.
For instance, the two CDW domains are made in this way in Fig.\ref{Circuit}%
(a). Applying $E_{z}$ only to a part in the QSH domain near the SQAH domain,
we can turn this part into the CDW domain as in Fig.\ref{Circuit}(b). We
have thus changed the form of circuit by a pair annihilation of two
Y-junction by applying $E_{z}$. This will open a way to topological
electronics.

I am very much grateful to N. Nagaosa for many fruitful discussions on the
subject. This work was supported in part by Grants-in-Aid for Scientific
Research from the Ministry of Education, Science, Sports and Culture No.
22740196.

\end{document}